\documentclass[journal]{IEEEtran}
\usepackage[utf8]{inputenc}
\usepackage[detect-weight=true, binary-units=true, per-mode=symbol]{siunitx}
\DeclareSIUnit \mhs {MHS06}
\usepackage{ifpdf}

\usepackage{cite}

\ifCLASSINFOpdf
  \usepackage[pdftex]{graphicx}
  \graphicspath{{figs/}}
  \DeclareGraphicsExtensions{.pdf,.jpeg,.png}
\else
  \usepackage[dvips]{graphicx}
  \graphicspath{{figs/}}
  \DeclareGraphicsExtensions{.eps}
\fi
\usepackage{array}
\usepackage{booktabs}

 \usepackage[caption=false,font=footnotesize]{subfig}
\begin{document}

\bstctlcite{IEEEexample:BSTcontrol}

%
\title{The CMS Data Acquisition System\\for the Phase-2 Upgrade}
%
%
%

\author{%
Jean-Marc~Andr\'{e}\IEEEauthorrefmark{5},
Ulf~Behrens\IEEEauthorrefmark{1},
Andrea~Bocci\IEEEauthorrefmark{2},
James~Branson\IEEEauthorrefmark{4},
Sergio~Cittolin\IEEEauthorrefmark{4},
Diego~Da~Silva~Gomes\IEEEauthorrefmark{2},
Georgiana-Lavinia~Darlea\IEEEauthorrefmark{6},
Christian~Deldicque\IEEEauthorrefmark{2},
Zeynep~Demiragli\IEEEauthorrefmark{6},
Marc~Dobson\IEEEauthorrefmark{2},
Nicolas~Doualot\IEEEauthorrefmark{5},
Samim~Erhan\IEEEauthorrefmark{3},
Jonathan~Richard~Fulcher\IEEEauthorrefmark{2},
Dominique~Gigi\IEEEauthorrefmark{2},
Maciej~Gladki\IEEEauthorrefmark{2},
Frank~Glege\IEEEauthorrefmark{2},
Guillelmo~Gomez-Ceballos\IEEEauthorrefmark{6},
Magnus~Hansen\IEEEauthorrefmark{2},
Jeroen~Hegeman\IEEEauthorrefmark{2},~\IEEEmembership{Member,~IEEE,}
Andr\'{e}~Holzner\IEEEauthorrefmark{4},
Michael~Lettrich\IEEEauthorrefmark{2},
Audrius~Mecionis\IEEEauthorrefmark{5}\IEEEauthorrefmark{9},
Frans~Meijers\IEEEauthorrefmark{2},
Emilio~Meschi\IEEEauthorrefmark{2},
Remigius~K.~Mommsen\IEEEauthorrefmark{5},
Srecko~Morovic\IEEEauthorrefmark{5}\IEEEauthorrefmark{10},
Vivian~O'Dell\IEEEauthorrefmark{5},
Samuel~Johan~Orn\IEEEauthorrefmark{2},
Luciano~Orsini\IEEEauthorrefmark{2},
Ioannis~Papakrivopoulos\IEEEauthorrefmark{7},
Christoph~Paus\IEEEauthorrefmark{6},
Andrea~Petrucci\IEEEauthorrefmark{8},
Marco~Pieri\IEEEauthorrefmark{4},
Dinyar~Rabady\IEEEauthorrefmark{2},
Attila~R\'{a}cz\IEEEauthorrefmark{2},
Valdas~Rapsevicius\IEEEauthorrefmark{5}\IEEEauthorrefmark{9},
Thomas~Reis\IEEEauthorrefmark{2},
Hannes~Sakulin\IEEEauthorrefmark{2}~\IEEEmembership{Member,~IEEE,}
Christoph~Schwick\IEEEauthorrefmark{2},
Dainius~\v{S}imelevi\v{c}ius\IEEEauthorrefmark{2}\IEEEauthorrefmark{9},
Mantas~Stankevicius\IEEEauthorrefmark{5}\IEEEauthorrefmark{9},
Jan~Troska\IEEEauthorrefmark{2},
Cristina~Vazquez~Velez\IEEEauthorrefmark{2},
Christian~Wernet\IEEEauthorrefmark{2},
Petr~Zejdl\IEEEauthorrefmark{5}\IEEEauthorrefmark{10}
\thanks{Manuscript received June 24, 2018.}
\thanks{This work was supported in part by the DOE and NSF (USA).}
\thanks{Corresponding author: Jeroen Hegeman (jeroen.hegeman@cern.ch)}
\thanks{\IEEEauthorrefmark{2}CERN,
  Geneva,
  Switzerland}%
\thanks{\IEEEauthorrefmark{1}DESY,
  Hamburg, Germany}%
\thanks{\IEEEauthorrefmark{5}FNAL, Batavia,
  Illinois,
  USA}%
\thanks{\IEEEauthorrefmark{6}Massachusetts
  Institute of Technology, Cambridge, Massachusetts,
  USA}%
\thanks{\IEEEauthorrefmark{3}University
  of California, Los Angeles, Los Angeles, California,
  USA}%
\thanks{\IEEEauthorrefmark{4}University of
  California, San Diego, San Diego, California,
  USA}%
\thanks{\IEEEauthorrefmark{7}Technical University of Athens, Athens,
  Greece}%
\thanks{\IEEEauthorrefmark{8}Rice University, Houston, Texas, USA}%
\thanks{\IEEEauthorrefmark{9}Also at Vilnius University,
  Vilnius, Lithuania}%
\thanks{\IEEEauthorrefmark{10}Also at CERN, Geneva,
  Switzerland}}

%
%

\markboth{544}%
{544}
%



\maketitle

\begin{abstract}
During the third long shutdown of the CERN Large Hadron Collider, the
CMS Detector will undergo a major upgrade to prepare for Phase-2 of
the CMS physics program, starting around 2026. The upgraded CMS
detector will be read out at an unprecedented data rate of up to
\SI{50}{\tera\bit\per\second} with an event rate of
\SI{750}{\kilo\hertz}, selected by the level-1 hardware trigger, and
an average event size of \SI{7.4}{\mega\byte}. Complete events will be
analyzed by the High-Level Trigger (HLT) using software algorithms
running on standard processing nodes, potentially augmented with
hardware accelerators. Selected events will be stored permanently at a
rate of up to \SI{7.5}{\kilo\hertz} for offline processing and
analysis. This paper presents the baseline design of the DAQ and HLT
systems for Phase-2, taking into account the projected evolution of
high speed network fabrics for event building and distribution, and
the anticipated performance of general purpose CPU. In addition, some
opportunities offered by reading out and processing parts of the
detector data at the full LHC bunch crossing rate
(\SI{40}{\mega\hertz}) are discussed.
\end{abstract}


%
\IEEEpeerreviewmaketitle

\section{Introduction}
\IEEEPARstart{T}{he} upgraded High-Luminosity LHC, after the third
Long Shutdown (LS3) will provide an instantaneous luminosity of
\SI{7.5e34}{\centi\meter^{-2}\second^{-1}} (levelled), with a pileup
of up to 200 interactions per bunch crossing.

Along with the increased statistics and thereby extended physics
reach, the increased HL-LHC luminosity carries several challenges. The
higher particle fluxes and total radiation doses require more
radiation-tolerant detectors and front-end electronics
technologies. While the increased instantaneous luminosity boosts the
probability of observing rare interactions, it at the same time
complicates triggering, reconstruction, and analysis of the event
data, due to the higher number of proton-proton collisions within the
same bunch crossing (the so-called `pileup').

The design goal for the CMS Phase-2 upgrade is: `to maintain the
current excellent performance in terms of efficiency, resolution, and
background rejection for all physics
objects'~\cite{Schmidt:2263093}. In order to achieve this, the main
focus points are:
\begin{itemize}
\item Improved tracker and muon detector coverage towards the forward
  regions, and increased granularity, aimed at preserving the
  performance of the CMS particle flow reconstruction
  algorithms~\cite{Sirunyan:2017ulk} at higher pileup and occupancy.
\item Inclusion of tracking information in the first level
  trigger~\cite{Collaboration:2272264}. Double-sided detector modules
  will correlate hit pairs into `stubs' with a coarse transverse
  momentum estimate. Only the data corresponding to tracks with
  $p_{\mathrm{T}} \geq \SI{2}{\giga\electronvolt}$ will be transmitted
  to the tracker back-end. This self-seeding approach results in an
  immediate tenfold rate reduction. The baseline design uses Hough
  transforms implemented in FPGAs to fit up to $2500$ tracks per bunch
  crossing, within the \SI{4}{\micro\second} latency allotment (out of
  a total trigger latency of \SI{12.5}{\micro\second}). The result is
  a significant improvement in the first-level trigger turn-on curves.
\item Addition of timing information to the front-end data of several
  subdetectors, in order to help with pileup suppression by adding
  coincidence information to spatial hits.
\item The introduction of a dedicated MIP\footnote{Minimum Ionizing
  Particle (MIP): a particle whose mean energy loss when passing
  through matter is close to the minimum.} timing
  detector~\cite{Collaboration:2296612}, closely integrated with the
  outer silicon tracker. Maintaining vertexing resolution at high
  pileup by spatial tracking improvements alone is extremely
  challenging. The combination of timing information with the tracker
  hits allows clustering, tracking and vertexing to take place in four
  dimensions instead of in just spatial 3D. Preliminary simulation
  indicates that a hit timing resolution of $O(\SI{30}{\pico\second}$)
  reduces the effective pileup from $O(200)$ to $O(40-50)$, the same
  level successfully handled by current CMS analyses.
\end{itemize}

More in-depth information on the CMS Phase-2 upgrade can be found in
the Technical Proposal~\cite{Contardo:2020886} and the various
Technical Design Reports (available on the CERN Document
server~\cite{cern:cds}).

\section{Overall design of the CMS Phase-2 DAQ system}
\begin{figure*}[!ht]
  \centering
  \includegraphics[width=.7\textwidth]{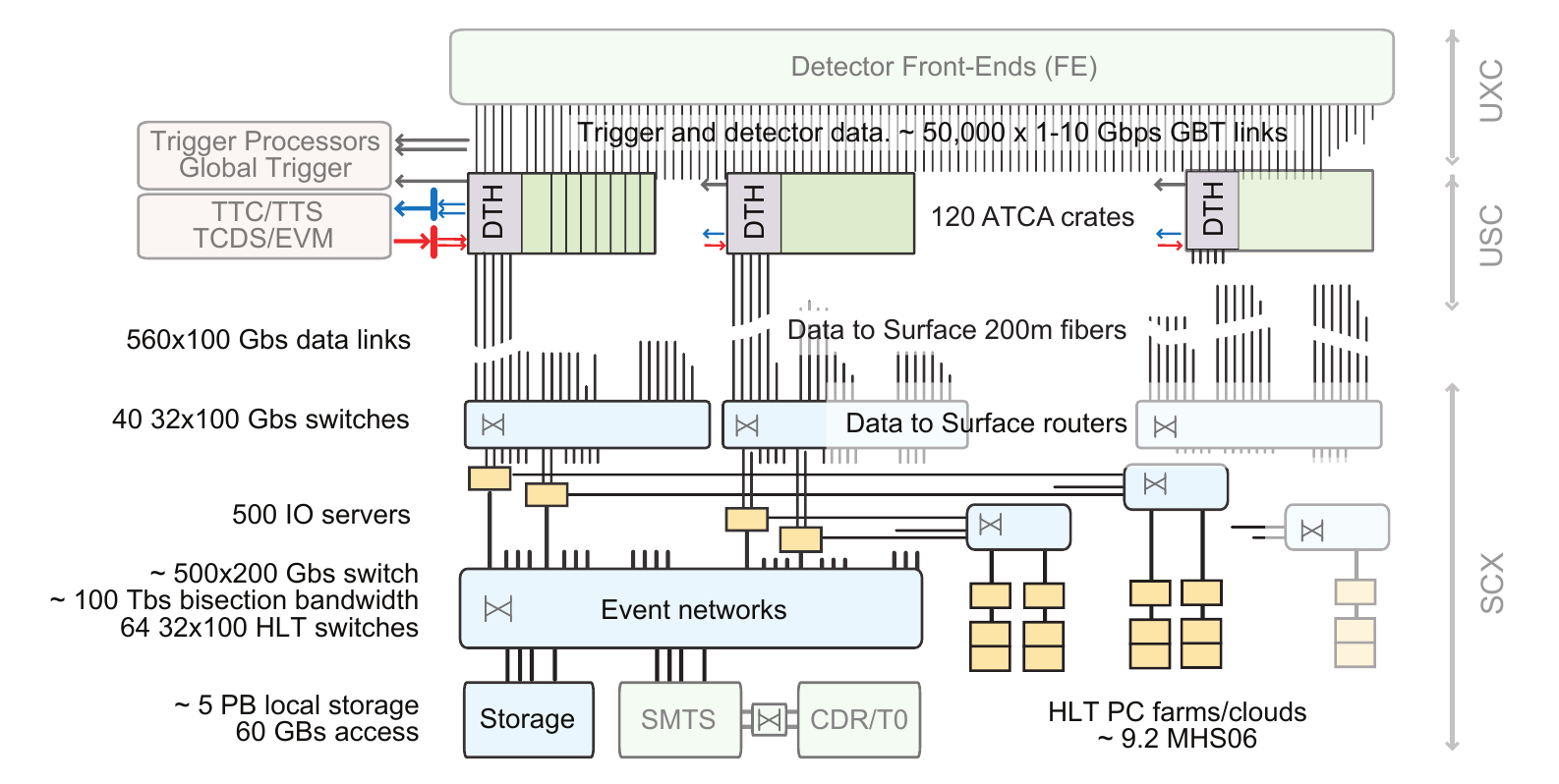}
  \caption{Overview of the CMS Phase-2 DAQ and Trigger system. The
    upgrade design still follows the original two-level design with a
    hardware trigger (Level-1) and a software trigger (High-Level
    Trigger). The main changes lie in the interconnect architecture
    and the integration of the DAQ and Timing Hub in the subsystem
    back-end crates.}
  \label{fig:baseline_overview}
\end{figure*}

Apart from the increased throughput and rate requirements of the
upgraded detector, the Phase-2 upgrade offers a unique moment of
reflection. Unlike the typical end-of-life replacements, or even the
Phase-1 upgrade\footnote{The CMS Phase-1 upgrade took place in the
  first long shutdown of the LHC (LS1, in 2013-2015), which prepared
  the accelerator and the experiments to operate at the higher
  center-of-mass energy of \SI{13}{\tera\electronvolt}.}, the CMS-wide
Phase-2 upgrade (due to its replacement of both front-end and back-end
electronics) allows breaking backward compatibility. As such, it is an
excellent moment to reassess current approaches and implementations.

The original CMS trigger-DAQ design~\cite{Cittolin:578006} combines a
hardware trigger (Level-1), implemented in custom electronics, with a
High-Level Trigger implemented in software and running on commodity
compute nodes. This design has proven to be very flexible and
scalable~\cite{Bauer:1457792,Andre:2016ibh}, and the Phase-2 baseline
DAQ design builds on this same architecture.

The upgrade of the Level-1 trigger falls outside the scope of this
paper, but is detailed in~\cite{Collaboration:2283192}. An interim
design report has been prepared for the Phase-2 CMS DAQ
upgrade~\cite{Collaboration:2283193}, and will be followed up by a
Technical Design Report for the DAQ and High-Level Trigger upgrade by
mid-2021. The remainder of this paper discusses the overall design and
scale of the Phase-2 CMS DAQ system, with a focus on open questions
and ongoing R\&D.

Figure~\ref{fig:baseline_overview} shows an overview of the baseline
Phase-2 DAQ design. The subdetector front-ends and back-ends are not
in the scope of the DAQ/HLT project, but in the context of the Phase-2
upgrade they present an interesting transition. All Phase-2 CMS
front-end-to-back-end communication uses the CERN
GBT~\cite{Moreira:1091474} and Versatile Link~\cite{Amaral:1265172} or
lpGBT~\cite{lpgbt} and Versatile Link+~\cite{vlplus} links, combining
data transport in one direction with clock and fast-control in the
other direction. This means that there no longer are any
timing-specific ASICs present in the front-ends, and that the central
timing and fast-control system only interfaces with the subdetector
back-end electronics (and no longer with any on-detector end-points).

Subdetector back-ends will transmit data on custom point-to-point
links at \SI{16} or \SI{25}{\giga\bit\per\second} (an evolution of the
CERN S-link~\cite{cern:slink} protocol) to a DAQ and Timing Hub in the
same crate, which concentrates and balances these data for
transmission to the surface counting room.

The data-to-surface (D2S) network will be based on commercially
available hardware, and use a standard protocol. A networked event
builder will assemble all back-end event fragments into events, and
transfer them to the High-Level Trigger filter farm. Events accepted
by the HLT are buffered locally in anticipation of transfer to the
CERN computing center for permanent storage.

At a pileup of 200 proton-proton collisions, the upgraded CMS detector
will produce an estimated event size of
$\approx\SI{7.4}{\mega\byte}$. At the design Level-1 accept rate of
\SI{750}{\kilo\hertz} this corresponds to an event-builder throughput
of \SI{44}{\tera\bit\per\second}. The HLT is supposed to reduce this
to an output rate of \SI{7.5}{\kilo\hertz} to storage.

\section{DAQ and Timing Hub}
\begin{figure}
  \centering
  \includegraphics[width=\columnwidth]{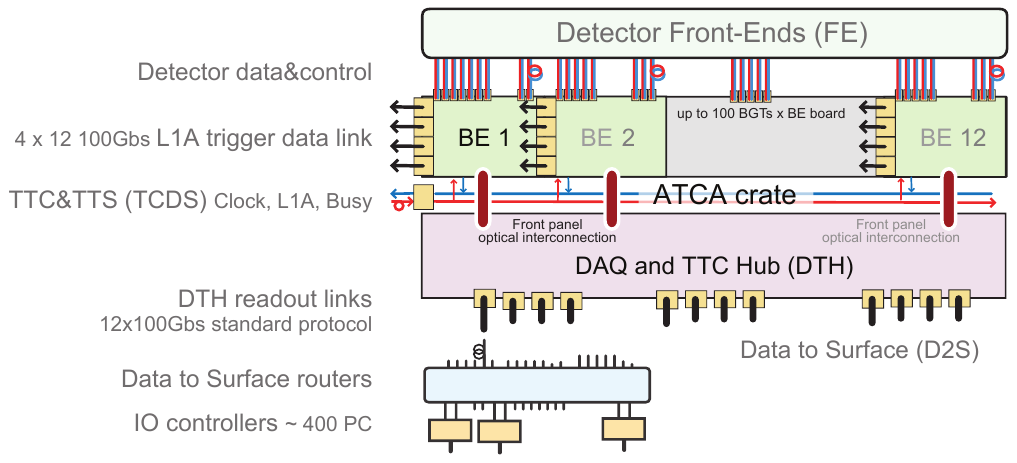}
  \caption{A DAQ and Timing Hub (DTH) in a back-end ATCA
    crate. Front-end to back-end connectivity is
    point-to-point. Level-1 trigger information is diverted by the
    back-end boards, and the data are sent to the DTH on
    point-to-point optical links. The DTH balances and concentrates
    these data, and sends them on to the commercial data-to-surface
    network.}
  \label{fig:dth_in_crate}
\end{figure}

One of the more fundamental changes of the Phase-2 DAQ upgrade is a
tighter integration among trigger control, fast control, data flow
monitoring, and the subdetector back-end electronics. The form factor
of choice for the Phase-2 CMS back-end and trigger-DAQ electronics is
ATCA~\cite{picmg:atca, specification:1159877}. Each back-end crate
will be equipped with a DAQ and Timing Hub (DTH). (See
Fig.~\ref{fig:dth_in_crate}.) This hub board has two central
functions: it interfaces the boards in the crate to the central Timing
and Trigger Control and Distribution System (TCDS), and it bridges the
custom back-end electronics to the standard data-to-surface
network. On the TCDS side the DTH also provides stand-alone
functionality for single-crate commissioning and testing. The DAQ side
will receive back-end data on custom point-to-point links (either
$24\times\SI{16}{\giga\bit\per\second}$ or
$16\times\SI{25}{\giga\bit\per\second}$) via the front panel and
concentrate and balance these data for transmission on the D2S network
(likely again via the front panel). The latter functionality also
requires large buffers to decouple the back-ends from any possible
network performance fluctuations.

A single DTH is designed to provide a DAQ throughput of
\SI{400}{\giga\bit\per\second} per crate (and is therefore often
called a `DTH400'), which is sufficient for most subdetectors. For
those subdetectors that require a higher throughput, a companion board
will be designed, the DAQ800, providing an additional throughput of
\SI{800}{\giga\bit\per\second} but no timing functionality.

The DTH is the main custom hardware deliverable for the Phase-2 DAQ
system, and plays a crucial role in the distribution of clock and
timing information from the central system to the back-end
electronics. A detailed prototyping program is under way, starting
with an `evaluation' prototype instrumented to prove timing
distribution performance in collaboration with back-end developers and
timing experts. The next steps include an updated prototype, produced
in moderate quantities, targeting test stands and firmware and/or
software development setups. The schedule leaves sufficient room for
an additional revision if necessary, either to fix oversights or to
catch up with advances in technology, before the production round of
the final hardware to be installed at the experiment.

\section{Data-to-Surface and event building networks}

\begin{table}[!t]
  \caption{Current estimate of the size of the Phase-2 data-to-surface
    network. Split by subdetector the diversity in throughput and
    subsystem size is clear. Accommodating this diversity
    cost-effectively in the central DAQ system is one of the
    challenges for the Phase-2 DAQ design.}
  \label{tab:daq4_d2s}
  \centering

  \begin{tabular}{lrrr}
    \toprule

    Sub-detector
    & back-end
    & Min.\ D2S
    & Min.\ D2S\\

    & crates
    & links/crate
    & links\\

    \midrule

    Outer tracker
    & 18
    & 4
    & 72\\

    Track trigger
    & 18
    & 1
    & 18\\

    Inner tracker
    & 4
    & 24
    & 96\\

    MIP timing detector - barrel
    & 1
    & 2
    & 2\\

    MIP timing detector - endcap
    & 1
    & 3
    & 3\\

    ECAL - barrel
    & 12
    & 9
    & 108\\

    HCAL - barrel
    & 2
    & 9
    & 18\\

    HCAL - HO
    & 1
    & 2
    & 2\\

    HCAL - HF
    & 1
    & 4
    & 4\\

    Endcap calorimeter - readout
    & 9
    & 18
    & 162\\

    Endcap calorimeter - L1 primitives
    & 12
    & 2
    & 24\\

    Muon - DT
    & 8
    & 1
    & 8\\

    Muon - CSC
    & 2
    & 6
    & 12\\

    Muon - GEM GE1/1
    & 1
    & 1
    & 1\\

    Muon - GEM GE2/1
    & 1
    & 1
    & 1\\

    Muon - GEM ME0
    & 1
    & 9
    & 9\\

    Muon - RPC
    & 1
    & 1
    & 1\\

    Level-1 trigger
    & 14
    & 1
    & 14\\

    Luminosity
    & 1
    & 1
    & 1\\

    \midrule

    Total
    &
    &
    & 556\\

    \bottomrule
  \end{tabular}

\end{table}

As was the case in previous generations of the CMS DAQ system, the
data-to-surface network will be based on commercial hardware and
standard protocols. The choice of the exact protocol and network
technology is still under consideration, and should become clear by
the time of the DAQ Technical Design Report in 2021. One important
constraint arises from the fact that the source of the D2S network is
the DTH custom hardware. The DTH data reception and data handling will
be implemented in an FPGA, suggesting a preference for a D2S protocol
that can be implemented in an FPGA as well. The CMS DAQ group has
experience with TCP/IP implementations in various FPGA families, and
at multiple speeds~\cite{Bauer:1639563, Gigi:2312401}. The baseline
D2S technology at the time of writing is
$4\times\SI{25}{\giga\bit\per\second}$ Ethernet $\rightarrow$
\SI{100}{\giga\bit\per\second} Ethernet using 100GBASE-CWDM4
interfaces to handle the approximately \SI{250}{\meter} cable length.

Table~\ref{tab:daq4_d2s} shows the estimated throughput requirements
on the D2S network for the different CMS subdetectors. Allowing for
headroom and non-optimal link use due to imperfect balancing based on
the separation into subdetectors, the estimated size of the D2S
network is $O(600)$ \SI{100}{\giga\bit\per\second} links. Implementing
this in a flexible and cost-effective fashion is one of the challenges
of the Phase-2 DAQ upgrade.

The D2S network brings the data to the surface counting room. There,
the baseline design foresees a set of commodity I/O servers handling
the assembly of event fragments into events. These I/O servers will be
interconnected by a dedicated, high-performance, event-building
network. Studies are ongoing to investigate the relative merits of a
traditional `linear' network layout, in which data flows
unidirectionally through the network, vs.\ a `folded' network
architecture. This `folded' architecture combines the data read-out
from the D2S network in the same hosts as the event building. In this
configuration each of these hosts serves both as source and as
destination for the event-building traffic, which flows through the
same network link. This makes the `folded' approach intrinsically more
efficient in terms of network usage. The `linear' approach, on the
other hand, benefits from performance tuning exploiting the
unidirectional data flow. The main (cost) difference, however, is
expected to lie in the number of required event-builder switch ports,
which can be significantly lower in a `folded' architecture.

\section{High-Level Trigger farm}

\begin{figure*}
  \centering
  \subfloat{\includegraphics[height=2.6in]{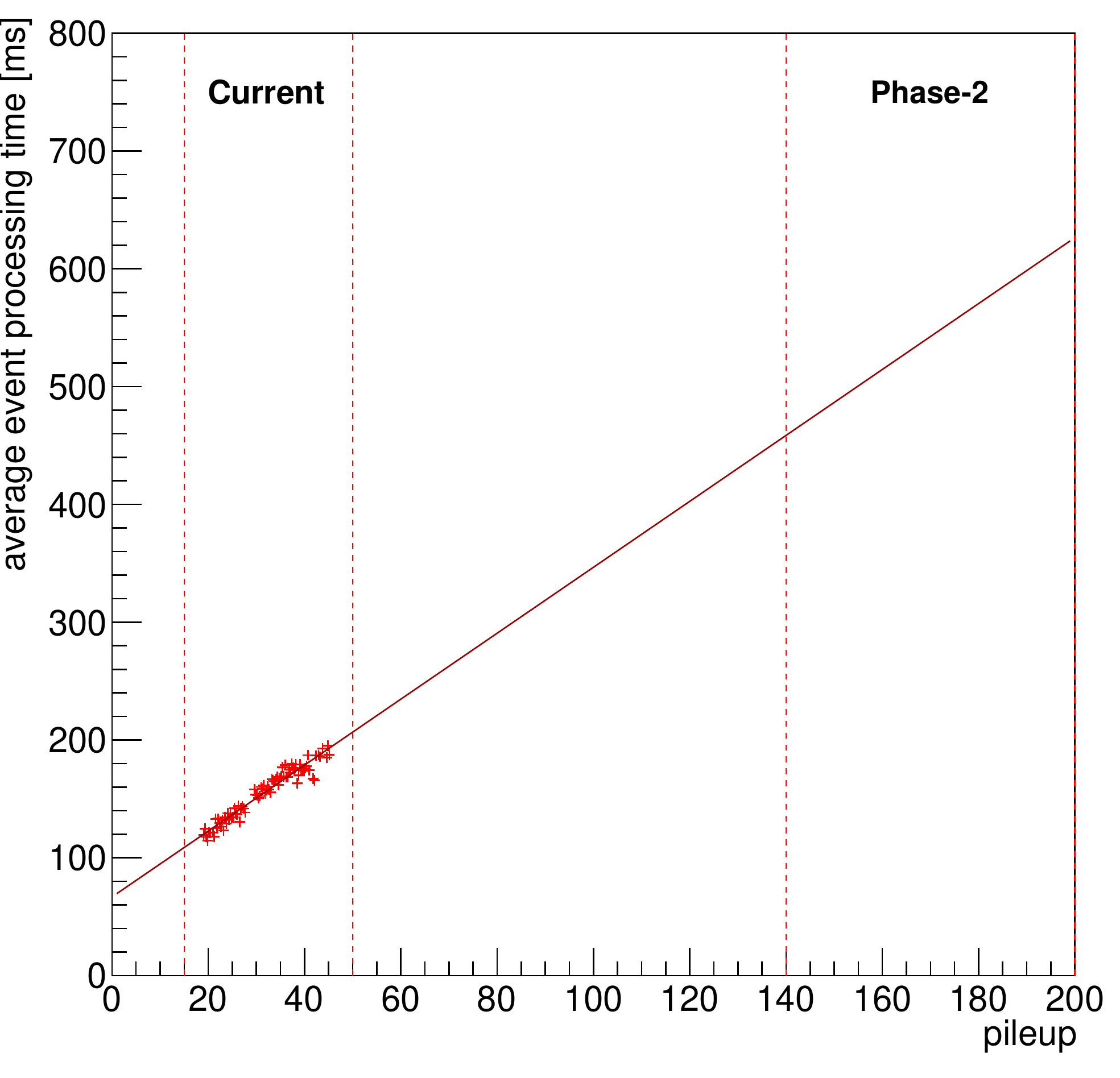}}
  \hfil
  \subfloat{\includegraphics[height=2.6in]{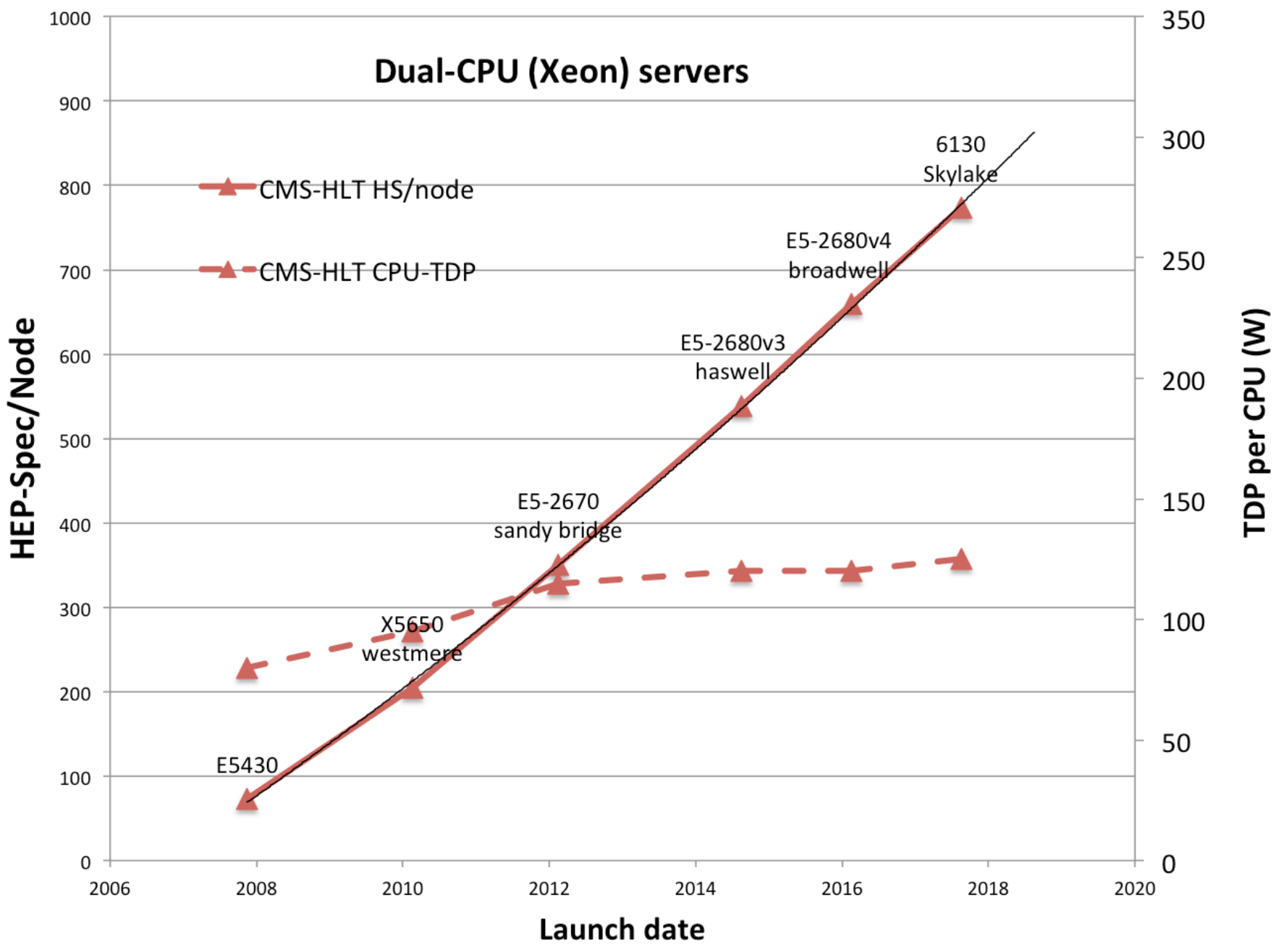}}
  \caption{Left: pileup dependence of the per-event HLT processing
    time required for the current run-2 detector. Naive extrapolation
    from the current (pileup $\approx 20$--$40$) to Phase-2 conditions
    (pileup $\approx 200$) would imply a quadrupling of the processing
    time. Right: evolution of compute power and thermal power per
    node, for the different generations of CPU used in the CMS
    HLT. The typical replacement cycle of the HLT compute nodes is
    five years.}
  \label{fig:hlt_projections}
\end{figure*}

Even with the addition of tracking information to the Level-1 trigger,
the Phase-2 operating conditions will require an increase in Level-1
output rate. Simulations show that, in order not to cut into the
physics signals, an increase from \SI{100}{\kilo\hertz} to
\SI{750}{\kilo\hertz} is required~\cite{Contardo:2020886}. This in
turn requires a significant increase in HLT compute capacity.

Figure~\ref{fig:hlt_projections} shows both the projected increase in
per-event compute time at a pileup of 200 and the evolution of compute
power and heat load per HLT node. Combined with the 7.5-fold rate
increase, it is clear that a more advanced solution than naive scaling
of the data center is necessary.

One possibility is the off-loading of compute-intensive tasks to
dedicated accelerator hardware, based, for example, on GPUs or
FPGAs. A preliminary study~\cite{Pantaleo:2293435} applying GPUs to
the pixel tracking shows a four-times gain in speed combined with a
$30\%$ reduction in power compared to the current CPU-only
implementation.

The current CMS R\&D program spans three possible strategies to add
accelerators to the HLT farm:
\begin{itemize}
\item Equipping all HLT compute nodes with co-processors. Of all
  approaches considered this is the most homogeneous solution, which
  at the same time makes it the hardest to optimize. For example:
  different tasks could benefit from different types of
  co-processors/accelerators. Homogeneous compute nodes would require
  multiplying the accelerator hardware throughout the whole HLT
  farm. In addition, this solution requires a careful balancing
  between the CPU and accelerator workloads in order to be profitable.
\item Creating a co-processor offload service on the event-builder
  network. This approach neatly sections off the special hardware in a
  `corner' of the event-builder network, and also trivially supports
  any admixture of different co-processors and/or accelerators. The
  downside, however, is that it requires additional high-bandwidth,
  low-latency connectivity between the HLT compute nodes and the
  co-processors. This latter point may affect the cost-benefit of this
  approach.
\item Integrating the co-processors directly in the I/O nodes
  receiving the data from the D2S network. While this is the most
  heterogeneous approach of all three, it has the benefit that it can
  be tailored easily by specializing the co-processors to the
  subdetector data being received. The challenge now lies in an
  effective separation of the accelerated pre-processing stage from
  the following HLT filtering stage. The most important downside of
  this approach is that the acceleration cannot be exploited for
  `on-demand' tasks, as it is not part of a consecutive filter chain
  like is the case in the HLT.
\end{itemize}

An important caveat applies from the point of view of cost: the
low-cost consumer-grade GPUs do not lend themselves for integration in
rack-based server infrastructure, forcing the choice of expensive GPUs
designed for data-center use.

While it is clear that naive scaling of the HLT farm will not be
sustainable for Phase-2, no clear candidate solution has been
pinpointed yet. One important open question is how much of the HLT
work load can be effectively ported to benefit from accelerators. The
road ahead depends not only on the outcome of the CMS studies, but
also on the evolution of technologies and prices in industry. By the
time of the Phase-2 DAQ TDR the ongoing R\&D is expected to enable the
formulation of a roadmap towards a final HLT design in time for
installation around 2026.

\section{Non-baseline studies}
The baseline Phase-2 DAQ design presented in the interim design
report~\cite{Collaboration:2283193} was carefully selected based on
its technical feasibility. In the absence of any external constraints
the system could be implemented with currently available
technologies. From the previous sections it will be clear that this
baseline design can (and has to) be improved in several respects.

In addition to investigations into improvements of the baseline
design, several studies are ongoing, that aim at providing additional
functionality and/or physics reach.

One interesting possible extension could be the addition of a parallel
`\SI{40}{\mega\hertz} scouting' read-out of all Level-1 trigger
primitives, augmented with the raw detector information where
bandwidth allows (e.g., the muon systems). Such a system would present
an `opportunistic experiment' in parallel to the CMS data-taking,
providing access to data from multiple successive bunch
crossings. Given the enormous data volumes involved, these data cannot
be stored permanently. Instead, a rolling window of data would be
kept, and analyzed using a combination of on-the-fly feature
extraction and query-based analyses. This approach could give access
to physics signatures that are not easily accessible with the
traditional hardware trigger approach. One example is the search for
rare, long-lived particles. Such studies would require simultaneous
access to detector information from multiple bunch crossings, which is
difficult to achieve and inefficient in a traditional hardware
trigger. Other physics cases include searches for very rare signatures
that are hidden by very high-rate backgrounds, such that it is
impossible to design a hardware trigger that is sufficiently selective
to reduce the trigger rate without affecting the signal. The analysis
of the scouting data could serve as `anomaly hunting', and lead to a
targeted trigger menu should any interesting signs of new physics be
found.

In addition to the physics potential, such a `\SI{40}{\mega\hertz}
scouting' system would provide an invaluable contribution in the form
of monitoring and calibration information with almost infinite
statistics.

\section{Summary and outlook}

\begin{table}[!h]
  \caption{The CMS Phase-2 DAQ system captured in numbers, and
    compared to the current Run-2 DAQ system.}
  \label{tab:daq4_in_numbers}
  \centering
  \begin{tabular}{lrrr}

    \toprule

    & \multicolumn{1}{r}{LHC} & \multicolumn{1}{r}{HL-LHC}\\
    CMS detector & \multicolumn{1}{r}{Run-2} & \multicolumn{1}{r}{Phase-2}\\

    Peak (avg.) pile-up
    & 60
    & 200\\

    \midrule

    Level-1 trigger accept rate (maximum)
    & \SI{100}{\kilo\hertz}
    & \SI{750}{\kilo\hertz}\\

    Event size
    & \SI{2.0}{\mega\byte}
    & \SI{7.4}{\mega\byte}\\

    Event network throughput
    & \SI{1.6}{\tera\bit\per\second}
    & \SI{44}{\tera\bit\per\second}\\

    Event network buffer (60 seconds)
    & \SI{12}{\tera\byte}
    & \SI{333}{\tera\byte}\\

    HLT accept rate
    & \SI{1}{\kilo\hertz}
    & \SI{7.5}{\kilo\hertz}\\

    HLT computing power
    & \SI{0.5}{\mhs}
    & \SI{9.2}{\mhs}\\

    Storage network throughput
    & \SI{2.5}{\giga\byte\per\second}
    & \SI{61}{\giga\byte\per\second}\\

    Storage capacity needed (1 day)
    & \SI{0.2}{\peta\byte}
    & \SI{5.3}{\peta\byte}\\

    \bottomrule
  \end{tabular}
\end{table}

Table~\ref{tab:daq4_in_numbers} shows a numerical comparison between
the current, Run-2, CMS DAQ system and the expected Phase-2 DAQ
system. From those numbers, and from the description in this paper, it
can be concluded that technologically the Phase-2 DAQ system could
already be realized at the time of writing. The real challenges lie in
the design of a cost-effective architecture that also satisfies the
financial, powering, and cooling constraints.

The prototyping program for the DAQ and Timing Hub should start
delivering answers within the next year, and the overall design of the
CMS Phase-2 DAQ and High-Level Trigger should become clearer by the
time of its Technical Design Report by mid-2021.



%








\bibliographystyle{IEEEtran}
\bibliography{IEEEabrv,rt2018_writeup}

\begin{thebibliography}{10}
\providecommand{\url}[1]{#1}
\csname url@samestyle\endcsname
\providecommand{\newblock}{\relax}
\providecommand{\bibinfo}[2]{#2}
\providecommand{\BIBentrySTDinterwordspacing}{\spaceskip=0pt\relax}
\providecommand{\BIBentryALTinterwordstretchfactor}{4}
\providecommand{\BIBentryALTinterwordspacing}{\spaceskip=\fontdimen2\font plus
\BIBentryALTinterwordstretchfactor\fontdimen3\font minus
  \fontdimen4\font\relax}
\providecommand{\BIBforeignlanguage}[2]{{%
\expandafter\ifx\csname l@#1\endcsname\relax
\typeout{** WARNING: IEEEtran.bst: No hyphenation pattern has been}%
\typeout{** loaded for the language `#1'. Using the pattern for}%
\typeout{** the default language instead.}%
\else
\language=\csname l@#1\endcsname
\fi
#2}}
\providecommand{\BIBdecl}{\relax}
\BIBdecl

\bibitem{Schmidt:2263093}
\BIBentryALTinterwordspacing
B.~Schmidt, ``{The High-Luminosity upgrade of the LHC: Physics and Technology
  Challenges for the Accelerator and the Experiments},'' \emph{J. Phys.: Conf.
  Ser.}, vol. 706, no.~2, p. 022002. 42 p, 2016. URL:
  \url{http://cds.cern.ch/record/2263093}
\BIBentrySTDinterwordspacing

\bibitem{Sirunyan:2017ulk}
A.~M. Sirunyan \emph{et~al.}, ``{Particle-flow reconstruction and global event
  description with the CMS detector},'' \emph{JINST}, vol.~12, no.~10, p.
  P10003, 2017.

\bibitem{Collaboration:2272264}
\BIBentryALTinterwordspacing
C.~Collaboration, ``{The Phase-2 Upgrade of the CMS Tracker},'' CERN, Geneva,
  Tech. Rep. CERN-LHCC-2017-009. CMS-TDR-014, Jun 2017. URL:
  \url{http://cds.cern.ch/record/2272264}
\BIBentrySTDinterwordspacing

\bibitem{Collaboration:2296612}
\BIBentryALTinterwordspacing
{CMS Collaboration}, ``{TECHNICAL PROPOSAL FOR A MIP TIMING DETECTOR IN THE CMS
  EXPERIMENT PHASE 2 UPGRADE},'' CERN, Geneva, Tech. Rep. CERN-LHCC-2017-027.
  LHCC-P-009, Dec 2017, this document describes a MIP timing detector for the
  Phase-2 upgrade of the CMS experiment, in view of HL-LHC running. URL:
  \url{https://cds.cern.ch/record/2296612}
\BIBentrySTDinterwordspacing

\bibitem{Contardo:2020886}
\BIBentryALTinterwordspacing
D.~Contardo \emph{et~al.}, ``{Technical Proposal for the Phase-II Upgrade of
  the CMS Detector},'' Geneva, Tech. Rep. CERN-LHCC-2015-010. LHCC-P-008.
  CMS-TDR-15-02, Jun 2015. URL: \url{http://cds.cern.ch/record/2020886}
\BIBentrySTDinterwordspacing

\bibitem{cern:cds}
\BIBentryALTinterwordspacing
 URL: \url{http://cds.cern.ch/}
\BIBentrySTDinterwordspacing

\bibitem{Cittolin:578006}
\BIBentryALTinterwordspacing
S.~Cittolin \emph{et~al.}, ``{CMS The TriDAS Project: Technical Design Report,
  Volume 2: Data Acquisition and High-Level Trigger. CMS trigger and
  data-acquisition project},'' Geneva, Tech. Rep., 2002. URL:
  \url{http://cds.cern.ch/record/578006}
\BIBentrySTDinterwordspacing

\bibitem{Bauer:1457792}
\BIBentryALTinterwordspacing
G.~Bauer \emph{et~al.}, ``{Operational experience with the CMS Data Acquisition
  System},'' CERN, Geneva, Tech. Rep. CMS-CR-2012-138, Jun 2012. URL:
  \url{http://cds.cern.ch/record/1457792}
\BIBentrySTDinterwordspacing

\bibitem{Andre:2016ibh}
J.-M. André \emph{et~al.}, ``{Performance of the new DAQ system of the CMS
  experiment for run-2},'' Tech. Rep., 2016.

\bibitem{Collaboration:2283192}
\BIBentryALTinterwordspacing
{CMS Collaboration}, ``{The Phase-2 Upgrade of the CMS L1 Trigger Interim
  Technical Design Report},'' CERN, Geneva, Tech. Rep. CERN-LHCC-2017-013.
  CMS-TDR-017, Sep 2017, this is the CMS Interim TDR devoted to the upgrade of
  the CMS L1 trigger in view of the HL-LHC running, as approved by the LHCC.
  URL: \url{http://cds.cern.ch/record/2283192}
\BIBentrySTDinterwordspacing

\bibitem{Collaboration:2283193}
\BIBentryALTinterwordspacing
{CMS Collaboration}, ``{The Phase-2 Upgrade of the CMS DAQ Interim Technical
  Design Report},'' CERN, Geneva, Tech. Rep. CERN-LHCC-2017-014. CMS-TDR-018,
  Sep 2017, this is the CMS Interim TDR devoted to the upgrade of the CMS DAQ
  in view of the HL-LHC running, as approved by the LHCC. URL:
  \url{http://cds.cern.ch/record/2283193}
\BIBentrySTDinterwordspacing

\bibitem{Moreira:1091474}
\BIBentryALTinterwordspacing
P.~Moreira \emph{et~al.}, ``{The GBT: A proposed architecure for multi-Gb/s
  data transmission in high energy physics},'' 2007. URL:
  \url{https://cds.cern.ch/record/1091474}
\BIBentrySTDinterwordspacing

\bibitem{Amaral:1265172}
\BIBentryALTinterwordspacing
L.~Amaral \emph{et~al.}, ``{The versatile link, a common project for
  super-LHC},'' \emph{JINST}, vol.~4, p. P12003, 2009. URL:
  \url{http://cds.cern.ch/record/1265172}
\BIBentrySTDinterwordspacing

\bibitem{lpgbt}
\BIBentryALTinterwordspacing
P.~Moreira, ``{The LpGBT Project, Status and Overview},'' March 2016, presented
  at the 2016 ACES workshop at CERN. URL:
  \url{https://indico.cern.ch/event/468486/contributions/1144369/}
\BIBentrySTDinterwordspacing

\bibitem{vlplus}
\BIBentryALTinterwordspacing
C.~Soos, ``{Versatile Link+},'' March 2016, presented at the 2016 ACES workshop
  at CERN. URL:
  \url{https://indico.cern.ch/event/468486/contributions/1144382/}
\BIBentrySTDinterwordspacing

\bibitem{cern:slink}
\BIBentryALTinterwordspacing
 URL: \url{http://hsi.web.cern.ch/HSI/s-link/}
\BIBentrySTDinterwordspacing

\bibitem{picmg:atca}
\BIBentryALTinterwordspacing
 URL: \url{https://www.picmg.org/openstandards/advancedtca/}
\BIBentrySTDinterwordspacing

\bibitem{specification:1159877}
\BIBentryALTinterwordspacing
``{Advanced TCA base specification: advanced TCA},'' Wakefield, MA, Tech. Rep.,
  2008. URL: \url{http://cds.cern.ch/record/1159877}
\BIBentrySTDinterwordspacing

\bibitem{Bauer:1639563}
\BIBentryALTinterwordspacing
G.~Bauer \emph{et~al.}, ``{10~Gbps TCP/IP streams from the FPGA for High Energy
  Physics},'' CERN, Geneva, Tech. Rep. CMS-CR-2013-402, Nov 2013. URL:
  \url{http://cds.cern.ch/record/1639563}
\BIBentrySTDinterwordspacing

\bibitem{Gigi:2312401}
\BIBentryALTinterwordspacing
D.~Gigi \emph{et~al.}, ``{The FEROL40, a microTCA card interfacing custom
  point-to-point links and standard TCP/IP},'' \emph{PoS}, vol. TWEPP-17, p.
  075. 5 p, 2017. URL: \url{http://cds.cern.ch/record/2312401}
\BIBentrySTDinterwordspacing

\bibitem{Pantaleo:2293435}
\BIBentryALTinterwordspacing
F.~Pantaleo \emph{et~al.}, ``{New Track Seeding Techniques for the CMS
  Experiment},'' 2017. URL: \url{https://cds.cern.ch/record/2293435}
\BIBentrySTDinterwordspacing

\end{thebibliography}
\end{document}